
\input phyzzx.tex
\def\a{\alpha}
\def\b{\beta}
\def\g{\gamma}

\def\d{\delta}

\def\t{\theta}

\def\r{\rho}

\def\k{\kappa}
\def\l{\lambda}

\def\s{\sigma}

\def\pa{\partial}
\def\na{\nabla}
\def\hg{\hat g}

\def\ov{\overline}
%
\def\ap#1{{\it Ann. Phys.} {\bf #1}}
\def\cmp#1{{\it Comm. Math. Phys.} {\bf #1}}

\def\pl#1{{\it Phys. Lett.} {\bf B#1}}
\def\prl#1{{\it Phys. Rev. Lett.} {\bf #1}}
\def\prd#1{{\it Phys. Rev.} {\bf D#1}}

\def\np#1{{\it Nucl. Phys.} {\bf B#1}}

\def\mpl#1{{\it Mod. Phys. Lett.} {\bf A#1}}

%
\REF\call{C.G. Callan, S.B. Giddings, J.A. Harvey, and A. Strominger,
\prd{45} (1992) R1005.}
\REF\hawk{S. W. Hawking,
\cmp{43} (1975) 199.}
\REF\sda{S. P. de Alwis, \pl{289} (1992) 278,\hfill\break
 hepth@xxx/9205069, \pl{} (in press); and Colorado preprint, (1992)
COLO-HEP-284, hep-th@xxx/9206020.}
 \REF\bilal{A. Bilal and C. Callan, Princeton preprint (1992)
PUPT-1320,\hfill\break hepth@xxx/9205089.}
\REF\banks{T. Banks, A. Dabholkar, M.R. Douglas, and M O'Loughlin,  Rutgers
preprint (1992) RU-91-54.}
\REF\lenny{J.G. Russo, L. Susskind, L. Thorlacius, Stanford preprint (1992)
SU-ITP-92-4, and L. Susskind and L. Thorlacius, Stanford preprint (1992)
SU-ITP-92-12. }
\REF\lyk{S. P. de Alwis and J. Lykken, \pl{269} (1991) 264.}
\REF\dkd{F. David, \mpl{3} (1988) 1651, J.Distler and H. Kawai, \np{321} (1989)
509.}
\REF\cham{A. Chamseddine,  \pl{256} (1991) 379, \pl{258} (1991) 97,and \np{368}
(1992) 98; J.M. Lichtzier and S.D. Odintsov, \mpl{6} (1991) 1953;
  A Chamseddine and Th. Burwick,
preprint (1992) hep-th@xxx/9204002;
J. Russo and A. Tseytlin, SU-ITP-2, DAMTP-1-1992. }
\REF\ham{K. Hamada, preprint (1992) UT-Komaba 92-7, hep-th@xxx/9206071; A.
Mikovic, preprint (1992) QMW/PH/92/12, hep-th@xxx/9207006.}
\REF\gsw{M. Green, J. Schwarz, E. Witten, Superstring Theory,  Chap. 3,
Cambridge University Press (1987).}
\REF\fri{D. Friedan, \prl{45} (1980) 1057.
C. Callan, D. Friedan,
E. Martinec, and M. Perry, \np{262} (1985) 593.
E.S. Fradkin and A. Tseytlin,
\pl{158} (1985) 316; \np{261} (1986) 413. S. Das and B. Sathiapalan, \prl{56}
(1986) 1511. R. Akhoury and Y. Okada, \pl{183}  (1987) 65.
C. Itoi and Y. Watabiki, \pl{198}  (1987) 486.}
\REF\as{A. Strominger, private communication.}
\REF\wit{E. Witten\np{311} (1988) 46.}
\REF\ct{T. Curtright and C. Thorn, \prl{48} (1982) 1309; E. Braaten, T.
Curtright, and C. Thorn, \pl{118} (1982) 115; \ap{147} (1983) 365; E. D'Hoker
and R. Jackiw, \prd{26} (1982) 3517.}
\REF\andy{A. Strominger ITP Santa Barbara preprint (1992)
UCSBTH-92-18,\hfill\break hepth@xxx/9205028.}
\REF\rst{J.G. Russo, L. Susskind, L. Thorlacius, Stanford preprint
SU-ITP-92-17.}
\REF\bd{N.D Birrell and P.C.W. Davies "Quantum Fields in Curved Space"
(Cambridge 1982) and references therein.}
\REF\gs{S. Giddings, and A. Strominger, Santa Barbara preprint (1992)
UCSBTH-92-28, hepth@xxx/9207034.}
\pubnum {COLO-HEP-288\cr hepth@xxx/9207095}
\date={July 1992; Revised October 1992}
\titlepage
\vglue .2in
\centerline{\bf Quantum Black Holes in Two Dimensions}
\author{ S.P. de Alwis\foot{dealwis@gopika.colorado.edu}}
\address{Dept. of Physics, Box 390,\break
University of Colorado,\break Boulder, CO
80309}
\vglue .2in
\centerline{\caps ABSTRACT}

We show that a whole class  of quantum actions for dilaton-gravity, which
reduce to the CGHS theory in the classical limit, can be written as a
Liouville-like theory. In a sub-class of this,  the field space singularity
observed by several authors is absent, regardless of the number of matter
fields, and in addition it is  such that  the dilaton-gravity functional
integration range  (the real line) transforms into itself for the Liouville
theory fields.  We also   discuss some problems
associated with the usual calculation of Hawking radiation, which stem from the
neglect of back reaction. We give an alternative argument incorporating back
reaction but find that the rate is still asymptotically constant. The latter is
due to the fact that the quantum
theory does not seem  to have a lower bound in energy and Hawking radiation
takes positive Bondi
(or ADM)  mass solutions to arbitrarily large negative mass.
\endpage

\chapter{\bf Introduction}
The theory of dilaton gravity coupled to scalar fields proposed by Callan et
al, [\call ] (CGHS) has generated a flurry of activity on black hole physics.
What one has is a simple toy model, within which the puzzling questions
associated with Hawking radiation [\hawk ] can be addressed in a systematic
way.
In the original work of  CGHS as well as in several subsequent papers, it was
assumed that quantum effects to leading order could be included by just adding
a piece to the action which reproduced the conformal anomaly. However it was
later realized that the consistent quantization of the theory in conformal
gauge, required that the cosmological constant term and/or the kinetic terms
should get   renormalized in a dilaton dependent manner, so that the theory
becomes a conformal field theory (cft)[\sda, \bilal ]. This requirement that
the theory be an exact cft (though not necessarily a soluble one) is not a
matter of choice. It is a  necessary consequence of general covariance. In
other words dilaton gravity coupled to matter fields
must be a cft in exactly the same way that string theory (i.e ordinary 2d
gravity coupled to matter fields) is a cft.

 In this paper we  will first review  this argument and then consider the
generalization of previous solutions to the conformal invariance conditions. We
show that there is a subset of models which  are free of the quantum black hole
singularity  pointed out in [\banks, \lenny], and which are such that the
original range of integration for the conformal factor and the dilaton, is
transformed into itself for the Liouville theory fields.  We will also argue
that  the calculations of Hawking radiation that have been given in the
literature, are inconsistent with the constraints and equations of motion of
the
theory since they
neglect back reaction. There is no sensible approximation scheme in which the
latter can be ignored.  We then  show that when the exact solution of the
system of equations coming from quantum corrected action is considered, the
results differ  from previous calculations. However it turns out that one
cannot see the radiation turning off in this theory, the (Bondi) mass of the
solutions of the
theory can be arbitrarily negative, and the  Hawking process causes a positive
mass solution to decay indefinitely to infinitely negative mass. Altough
Liouvile theory has a positive definite spectrum   the same is not true of the
Liouville-like theory that is obtained from the CGHS theory. It is possible
 that the origin of the problem lies in the  the CGHS theory itself, but a more
rigorous quantum treatment of the Bondi mass may resolve this question.

In the next section we review the quantization of the CGHS theory. In the third
section we discuss a class of  solutions to the integrability conditions for
the constraints and present arguments for taking the resulting exact conformal
field theory as a quantum theory of dilaton gravity. In the fourth section we
demonstrate explicitly how the classical singularities are tamed by quantum
effects. In the fifth section we
review the CGHS calculation of Hawking radiation in this model. In the sixth
section we give an alternative calculation which is consistent with the
constraints (this is basically a detailed version of a calculation contained in
the second paper of [\sda ]) and in the final section we make some concluding
remarks.
\chapter{\bf Quantization}
The CGHS theory is defined by the classical action

$$S={1\over 4\pi}\int d^2\s\sqrt{-g}[e^{-2\phi}(R+4(\na\phi )^2+4\l^2)-{1\over
2}\sum_{i=1}^N (\na f^i)^2 ].\eqn\cghs$$

 In the above $G$ is the $2d$ metric, $R$ is its curvature scalar, $\phi $ is
the dilaton and the $f^i$ are $N$ scalar matter fields. This action may be
obtained as a  low energy effective action  from string theory, in which case
 the $f$ fields will arise from the Ramond-Ramond sector. Note that the (zero
mass) tachyon of $2d$ string theory is excluded from this action. If this field
had been coupled then the theory would not be solvable even at the classical
level.\foot{For a discussion of how in this case, $2d$ black hole solutions are
affected far away from the black hole by the presence of the tachyon, see [\lyk
].}

The quantum field theory of this classical action may be defined as

$$Z=\int {[dg]_g[d\phi ]_g[df]_g\over [Vol.~Diff.]} e^{iS[g,\phi,f]}.\eqn\qft$$

The metrics which define these measures are usually given by,

$$\eqalign{&||\d g||_g^2=\int d^2\s\sqrt{- g}g^{\a\g}g^{\b\d}(\d g_{\a\b}\d
g_{\g\d}
+\d g_{\a\g}\d g_{\b\d} )\cr
||\d\phi ||_g^2 &=\int d^2\s\sqrt{- g}\d\phi^2,\quad ||\d f||_g^2=\int
 d^2\s\sqrt{- g}\d_{ij}\d f^i\d f^j.\cr}\eqn\met$$
  However we can be more general  in these definitions as long as $2d$
diffeomorphism invariance is preserved. Now
  let us gauge fix to the conformal gauge $g=e^{2\r}\hg$ and rewrite the
measures with respect to the fiducial metric $\hg$. Following the work of
  David and of Distler and Kawai [\dkd ], we may expect the action to get
renormalized, except that unlike in their case the renormalization will be
  dilaton dependent (since the coupling is $e^{2\phi})$. Thus in general we may
expect the gauge fixed path integral to be written as [\cham,\lyk,\sda]\foot
{For alternative approaches to the quantization  see [\ham ].},

  $$Z=\int [dX^{\mu}]_{\hg}[df]_{\hg}([db][dc])_{\hg}e^{iI(X,\hg)+iS(f,\hg )+
 iS(b,c,\hg )},\eqn\part$$
  where
  $$I[X,\hg ]=-{1\over 4\pi}\int \sqrt{-\hg}[{1\over 2}
  \hg^{ab}G_{\mu\nu}\pa_aX^{\mu}\pa_bX^{\nu}+\hat R\Phi (X)
+T(X)].\eqn\sigmod$$
$S(b,c,\hg )$ is the Fadeev-Popov ghost action, and we have written $(\phi ,\r
)=X^{\mu}$.
Note that all the measures in \part\ are defined with respect to the $2d$
metric $\hg$ and that in particular  the measure $[dX^{\mu}]$ is derived from
the natural metric on
the space $||\d X_{\mu}||^2=\int d^2\s \sqrt{-\hg}G_{\mu\nu}\d X^{\mu}\d
X^{\nu}$.
 {\it In the limit of weak coupling} ($e^{2\phi}<<1$) we have,

 $$\eqalign {I\rightarrow &{1\over 4\pi}\int d^2\s\sqrt{-\hg}[e^{-2\phi}
(4(\hat\na\phi )^2-4\hat\na\phi .\hat\na\r )-\k\hat\na\r .\hat\na\r\cr
&+\hat R(e^{-2\phi}-\k\r )-4\l^2e^{2(\r-\phi )}]\cr}\eqn\cacn$$

This is obtained from \cghs\ by putting $g=e^{2\r}\hg$, and including a very
specific higher order term; namely the usual conformal anomaly term. $\k$ in
the above is equal to ${26-(N+2)\over 6}={24-N\over 6},$ if one includes the
contribution of the transformation of the measure for $\phi$ and
$\r$.\foot{We will justify this  in more detail later on.}

$I$ is a generalized sigma model action and we have kept only renormalizable
terms. The sigma model action  introduces three (dilaton dependent) coupling
functions
$G,\Phi$, and $T$, respectively the field space metric, dilaton, and tachyon.
The only a priori restriction arises from the fact that the functional integral
for $Z$ in \part, must be independent of the fiducial metric $\hg$, as is
obvious from
the expression \qft\ for it. This implies that the following constraints should
be satisfied:
$$<T_{\pm\pm}+t_{\pm\pm}>=0,\eqn\cons$$
and
$$<T_{+-}+t_{+-}>=0,\eqn\conf$$
where $T_{\mu\nu}$ is the stress tensor for the dilaton-gravity and matter
sectors, and $t_{\mu\nu}$ is the stress tensor for the ghost sector. (\conf\ is
equivalent to the equation of motion for $\r$, and so is not an additional
constraint). Furthermore one has to satisfy the integrability conditions for
these constraints, namely that they generate a Virasoro algebra with zero
central charge.\foot{In effect this means that the field space must be  exactly
like the target space of string theory, though here we do not give this space a
space-time interpretation. The only space-time in the theory is the original
one parametrized by the coordinates $\s$.} As is well known (see for instance
[\gsw ] and references therein) this is equivalent to the requirement that the
$\b$-functions [\fri ] corresponding to the coupling functions, $G,\Phi$, and
$T$, vanish.

 $$\eqalign{\b_{\mu\nu}&={\cal R}_{\mu\nu} +2\na_{\mu}^G\pa_{\nu}\Phi-
 \pa_{\mu}T\pa_{\nu}T+\ldots ,\cr
 \b_{\Phi}&=-{\cal R}+4G^{\mu\nu}\pa_{\mu}\Phi\pa_{\nu}\Phi-4\na_G^2\Phi+
 {(N+2)-26\over 3}+G^{\mu\nu}\pa_{\mu}T\pa_{\nu}T-2T^2+\ldots ,\cr
 \b_T&=-2\na^2_GT+4G^{\mu\nu}\pa_{\mu}\Phi\pa_{\nu}T-4T+\ldots ,\cr}\eqn\bet$$

 where {\cal R} is the curvature of the metric $G$. These equations have to be
solved under the boundary conditions that in the weak coupling limit
($e^{2\phi}<<1$) we get, comparing \sigmod\ with \cacn ,

 $$G_{\phi\phi}=-8e^{-2\phi},\quad G_{\phi\r}=4e^{-2\phi},\quad
G_{\r\r}=2\k,\eqn\metr$$
 $$\Phi =-e^{-2\phi}+\k\r ,\quad T=-4\l^2e^{2(\r-\phi)}.\eqn\dilt$$

\chapter{\bf From CGHS to Liouville}
 Let us first discuss the renormalization of the field space metric and dilaton
($G$ and $\Phi$) and postpone the discussion of the tachyon $T$. The
(renormalized) field space metric may be parametrized as,

$$ ds^2=-8e^{-2\phi}(1+h(\phi ))d\phi^2+8e^{-2\phi}(1+\ov h(\phi ))d\r d\phi
 +2\k (1+\ov{\ov h})d\r^2 ,\eqn\metric$$
 where $h,\ov h,$ and $\ov{\ov h}$ are $O(e^{2\phi})$. If we are going to
consider only $O(e^{2\phi} )$ effects then we should certainly set $\ov{\ov h}$
to zero. But even if we consider the renormalization functions $h$ and $\ov h$
to all orders, it is consistent to limit ourselves to the class of quantum
versions of the CGHS theory which have $\ov{\ov h}=0$, provided that we satisfy
the beta function equations. This corresponds to confining ourselves to
theories in which the field space curvature ${\cal R}=0$. In this case we can
transform this metric to Minkowski form. First put\foot{We will only consider
theories with $\k\ne 0$ i.e. $N\ne 24$.}

 $$y=\r-\k^{-1}e^{-2\phi}+{2\over\k}\int d\phi e^{-2\phi}\ov h (\phi
).\eqn\ycdt$$

Then the metric becomes

$$ds^2 = -{8\over\k}P^2(\phi )d\phi^2+2\k dy^2,$$
where

$$P(\phi )=e^{-2\phi}[(1+\ov h)^2+\k e^{2\phi}(1+h)]^{1\over 2}.\eqn\pee$$

Putting

$$x=\int d\phi P(\phi ),\eqn\xcdt$$
 we have

$$ds^2=-{8\over\k}dx^2+2\k dy^2\eqn\newmet$$

With this form of the metric, ignoring $O(T^2)$ terms, we find from the first
(graviton) $\b$-function equation in \bet , that $\pa_{\mu}\pa_{\nu}\Phi =0$.
In other words
$\Phi$ is linear in $x,y$. Demanding that we recover the CGHS $\Phi$ given in
\dilt\ in the weak coupling limit we find the unique solution,

$$\Phi=\k y.\eqn\dilaton$$

Substituting in the second (dilaton) equation in \bet , we then get

$$\k ={24-N\over 6}.\eqn\kap$$

To determine $T$ we consider the third equation of \bet , to linear order and
get,

$${\k\over 4}\pa^2_xT-{1\over k}\pa_y^2T+2\pa_yT-4T=0.\eqn\taceq$$

This has solutions of the form $T=e^{\b x+\a y}$, where ${\k\over
4}\b^2-{1\over k}\a^2+2\a-4=0$. Now we need to impose the boundary condition
that we recover the CGHS tachyon given in \dilt\ in the weak coupling limit. To
do so we expand the expression for $x$ (\xcdt , \pee ) to get
$$x\simeq -{1\over 2}e^{-2\phi}+\int d\phi e^{-2\phi}\ov h+{\k\over
2}\phi+O(e^{2\phi}).\eqn\xweak$$
Then we find that
$-{4\over\k}x+2y=2\r-2\phi+O(e^{2\phi})$
so that the unique solution (confining ourselves to multiplicative
renormalizations) obeying the required boundary condition is
$$T=-4\l^2e^{-{4\over\k}x+2y}.\eqn\tachyon$$

For  $\k>0$ there is another (additive) term \foot{I wish to thank Andy
Strominger for pointing  this out to me [\as ].}  satisfying the boundary
condition. Namely
$$ T_{np}=\mu e^{4\over\sqrt{\k}x}\simeq\mu \exp
(-{2\over\sqrt\k}e^{-2\phi}).$$

This is in fact a non-perturbative ambiguity. We will set $\mu =0$ in the rest
of the paper. In any case it is absent for $\k <0$, since in that case we will
have an oscillatory solution which will not vannish in the classical limit.

It is convenient now to introduce rescaled fields,

$$X=2\sqrt{2\over |\k |}x,~~Y=\sqrt{2|\k|}y,\eqn\rescale$$

in terms of which the metric and the tachyon become,
 $$\eqalign{ds^2=&\mp dX^2\pm dY^2\cr
            T=&-4\l^2e^{\mp\sqrt{2\over |\k |}(X\mp Y)}\cr}$$
In the above and in the  equations below, upper/lower signs correspond to
having $\k >0$/$\k <0$ respectively.
In terms of the new field variables the functional integral becomes,

$$Z=\int [dX][dY][df][db][dc]e^{iS[ X,Y,f]+iS_{ghost}},\eqn\newqft$$
 where,

 $$S={1\over 4\pi}\int d^2\s[\mp\pa_{+}X\pa_{-}
X\pm\pa_{+}Y\pa_{-}Y+\sum_i\pa_{+}f^i\pa_{-}f^i+2\l^2e^{\mp\sqrt{2\over |\k
|}(X\mp Y)}].
\eqn\newaction$$

Several comments need to be made about this functional integral. First and most
obviously there is the question of the range of the integration. As we see from
\xcdt\ and \pee, in general the range of integration in $X$ will not extend
over the whole real line. What we then have is an approximate solution to the
$\b$-function equations \bet\
valid only to leading order in the sigma model ($\a '$) expansion and to
leading order in the weak field expansion in $T$. On the other hand if we {\it
define} the quantum theory by \newqft\ with the range of integration for $X$
being the whole real line, we have\foot{The theory is very much like Liouville
theory which is an exact cft [\ct ]. In fact it is less singular than
Liouville. So one  expects it to be an exact cft as well.}  a solution to the
exact $\b$-function equations. Thus this  definition of quantum dilaton-gravity
theory, even if  somewhat unorthodox, is a very compelling one. It is on the
same footing as for instance the definition of
$2+1$ dimensional quantum gravity given by Witten[\wit ] in which the
functional integral is taken over all values of the vielbein field.  Also  let
us point out that if we restrict the range of integration to be consistent with
the original  definition of the quantum theory then, since we only have a
solution to the leading order $\b$-function equations, it seems as if we will
need an infinite number of terms to  satisfy the exact conformal invariance
conditions. It is plausible to suppose that this theory is equivalent to the
one above with the unrestricted range of integration. This argument is also
reinforced by the fact that, as in the
usual Liouville theory, the integration range is effectively cut off (albeit
softly) by the Liouville potential term.
Finally (and perhaps this is the most compelling reason for the  quantum
Liouville-like conformal field theory) there exist  choices of $h$ and $\bar
h$,
for which when the integration ranges  for $\phi$ and $\r$ are as usual taken
over the whole real line,  the same is true for the ranges for $X$ and $Y$ (see
 case d at the end of this section).

The second comment is with regard to the approximation in which the dilaton and
graviton loops can be ignored. By rescaling and translating the fields $X,Y$ it
is easily seen that $\hbar =\k$ so that the semiclassical approximation is
valid only for large $\k$. Thus one might be inclined to believe that any (even
qualitative) conclusions derived for the $N<24$ theory[\sda ] are drastically
effected by dilaton graviton loop corrections. On the other hand for $N=1$ we
get $\k =3.8$  which is of the same order as the relevant parameter in QCD
where the approximation  works quite well.

Finally we comment on the different possibilities for the functions
$h$ and $\ov h$. Three special cases have so far been discussed in the
literature.

a) $h=\ov h=0$. i.e. the field-space metric of the classical CGHS Lagrangian is
not renormalized. However in this case the cosmological constant term $T$ is
renormalized.

b)$h=-e^{2\phi},~ \ov h= -2e^{2\phi}$. This is the case proposed by Strominger
[\andy ]. In this case both the metric $G$ and the tachyon $T$ are
renormalized.

c) $h=0,~\ov h=-{\k\over 4}e^{4\phi}$. This is the  case considered in [\rst ]
where $P^2$ is a perfect square (see \pee\ from which we find
$P=e^{-2\phi}(1+{\k\over 4}e^{2\phi}))$. In this case the metric is (obviously)
renormalized but the tachyon is not (as is easily seen from \tachyon\ and the
expressions for $x$ and $y$ with the above value of $P$).

d) In all of the above cases the transformation \xcdt\ has a singularity when
$\k <0$. For instance in case a) it is at $e^{2\phi}=-\k^{-1}$. It is however
quite easy to find a class  of models which have no such singularity. Put
$\ov h=ae^{2\phi}$ and $h=be^{2\phi}$. Then putting $e^{2\phi}=z$ the condition
for the absence of a singularity is that the quadratic equation $z^2P^2=(a^2+\k
b)z^2+(2a+\k)z+1=0$ has no real roots. i.e. we must choose
$\k^2+4(a-b)\k <0$. Obviously there are many solutions to these conditions but
one particular class is of particular importance since members of it naturally
allow the range of integration in the $X,Y$ variables to go over the whole real
line. The simplest member of this class has $h=0$ and $\bar h=-{\k\over
2}e^{2\phi}$. In this case we have from \pee,\xcdt,\ycdt,
$$\eqalign{x&=\int d\phi e^{-2\phi}(1+{\k^2\over 4}e^{4\phi})^{1\over 2}\cr
            &=-{1\over 2}({\k^2\over 4}+e^{-4\phi})^{1\over 2}+{|\k |\over
4}\sinh^{-1}\left ({|\k |\over 2}e^{2\phi}\right ),\cr}\eqn\newx$$
	    and
$$y=\r-\k^{-1}e^{-2\phi}-\phi .\eqn\newy$$

Clearly as $\phi ,\r$, range from $-\infty$ to $+\infty$
	so do $x$ and $y$.

\chapter{\bf Exact solutions}

 The equations of motion coming from \newaction\ are as follows.\foot{In this
section and in section 6, wherever it is appropriate, all equations are to be
understood as being valid inside the functional integral, i.e. as expectation
values of quantum operators. Since following the arguments of reference [15]
the theory can be mapped into a free theory it is plausible that the only
quantum effects come from normal ordering.}
 $$\pa_+\pa_-f=0,\eqn\feqn$$
$$\eqalign{\pa_+\pa_-X=&\l^2\sqrt{2\over |\k |}e^{\sqrt{2\over |\k |}(X+Y)},\cr
           \pa_+\pa_-Y=&-\l^2\sqrt{2\over |\k |}
	   e^{\sqrt{2\over |\k |}(X+Y)}.\cr}\eqn\xy$$

We have taken the case with the  lower signs in \newaction\ so that the
discussion is for  $N>24$. There is no qualitative difference in the other case
so it is unnecessary to write it out explicitly.\foot {It is also contained in
[\bilal ] and the second paper of [\sda ].} These equations are easily solved.
{}From \xy\ we have
$\pa_+\pa_-(X+Y)=0,$ so that $X+Y=\sqrt{|\k |\over 2}(g_+(\s^+)+g_-(\s^-)),$
where $g_{\pm}$ are arbitrary chiral functions. Substituting into the $X$
equation of motion  and integrating we have

 $$\eqalign{X=&-\sqrt{2\over |\k |}(u_+(\s^+)+u_-(\s^-))+\l^2\sqrt{2\over |\k
|}
\int^{\s^+}d\s^+e^{g_+(\s^+)}\int^{\s^-}d\s^-e^{g_-(\s^-)}\cr
 &=-Y+\sqrt{|\k |\over 2}(g_++g_-),\cr}\eqn\soln$$
 where $u_{\pm}$ are arbitrary chiral functions to be determined by the
boundary conditions.

 By a coordinate choice we can set $g_{\pm}=0$. In these coordinates (the
analog of Kruskal-Szekeres coordinates for the black hole) we get

 $$X=-Y=-\sqrt{2\over |\k |}(u-\l^2\s^+\s^-).\eqn\ssoln$$
where $u=u_++u_-$.

 These solutions are of course the same as those of  CGHS, except that they are
for $X$ and $Y$ and all the effects of the quantum anomalies are now
incorporated in the expressions for them in terms of $\r$ and $\phi$. To be
explicit consider the case d) discussed at the end of the last section
($h=0,~\bar h=-{\k\over 2} e^{2\phi}$);

 $$\eqalign{X=&2\sqrt{2\over |\k |}\int d\phi e^{-2\phi}[1+{\k^2\over
4}e^{4\phi}]^{1\over 2}\cr
 =&\sqrt{2|\k |}\int d\phi [1+{4\over\k^2}e^{-4\phi}]^{1\over 2},\cr}$$
 and
 $$Y=\sqrt{2|\k |}\r+\sqrt{2\over |\k |}e^{-2\phi}-\sqrt{2|\k |}\phi .$$

 In the weak coupling limit ($e^{2\phi} <<1$) we have from \ssoln\ the
classical solution

 $$e^{-2\phi}=e^{-2\r}=u-\l^2\s^+\s^-,\eqn\cghssoln$$

which exhibits the classical (black hole type) singularity on the curve where
the right hand side vanishes. But  the singularity  is in the strong coupling
region where we have to use the strong coupling expansion (from the second line
of the above equation  for $X$)
$$X\simeq\sqrt{2|\k |}[\phi -{e^{-4\phi}\over\k^2}+\ldots].$$

Then we have from  \ssoln,

$$\phi\simeq\k^{-1}(u-\l^2\s^+\s^-),$$
and
$$\r\simeq{1\over \k}e^{-2\k^{-1}(u-\l^2\s^+\s^-)}.$$

The metric ($e^{2\r}$) is clearly non-singular at the classical singularity.

 Differentiating the solution for $X$ with respect to $\s_{\pm}$ we get

 $$2e^{-2\phi}\pa_{\pm}\phi=-{(\pa_{\pm}u_{\pm}-\l^2\s^{\mp})\over \bar P(\phi
)},\eqn\dphi$$

where $\bar P = e^{2\phi}P$, $P$ being defined by \pee. This equation gives the
trajectory of the apparent horizon ($\pa_+\phi =0$) introduced in [\lenny ]
(once the unknown  function $u$ is determined) as

$$\s^-={1\over\l^2}\pa_+u_+(\s^+).\eqn\hor$$

By differentiating the solution for $Y$ and using \dphi\ and the expression
for $Y$ in terms of $\r ,~\phi$, we have

$$\eqalign{\k\pa_-\pa_+\r=(1+\bar h)\pa_-\bar h(\pa_+u_+-\l^2\s^-)\bar
P^{-1}&-(1+\bar h)(\pa_+u_+-\l^2\s^-)\pa_-\bar P\bar P^{-2}\cr&+\l^2\left
(1-{1+\bar h\over\bar P}\right ).\cr}\eqn\curv$$

 From this expression the curvature $R=8e^{-2\r}\pa_+\pa_+\r$ is easily seen to
be non-singular at the classical singularity in all the cases a) to d)
discussed at the end of section 3 (as is obvious from the fact that the metric
is non-singular there) and furthermore in case d), it is seen that there are
no curvature  singularities anywhere for either sign of $\k$.

\chapter{\bf Problems in Calculating  Hawking Radiation}

Before we calculate Hawking radiation we would like to comment on previous
calculations of this phenomenon in 2d dilaton gravity. These comments may  have
a bearing on the original calculation [\hawk ] in 4d as well.

In the CGHS calculation [\call ], the stress tensor anomaly is added to the
classical
stress tensor trace to give

$$T_{-+}=e^{-2\phi}(2\pa_+\pa_-\phi-4\pa_+\phi\pa_-\phi
)-\l^2e^{2\r-2\phi}-{N\over 6}\pa_+\pa_-\r.$$

{}From the conservation equation for the stress tensor the remaining
components of the stress tensor are then determined to be

$$T_{\pm\pm}=e^{-2\phi}(4\pa_{\pm}\r\pa_{\pm}\phi-2\pa^2_{\pm}\phi
)+T_{\pm\pm}^f,$$
 with the quantum (one loop) part of the stress tensor being given by

$$ T_{\pm\pm}^f=-{N\over 6}
(\pa_{\pm}\r\pa_{\pm}\r-\pa_{\pm}^2\r+t_{\pm}(\s^{\pm})),$$

 where $t_{\pm}$ are arbitrary chiral functions to be determined by the
 boundary conditions. Of course in a consistent quantization ghosts have to be
included and $N\rightarrow N-24$ [\andy ,\sda ,\bilal ] and $t_{\pm}$ must be
related to the ghost stress tensor [\sda , \bilal ] but we will ignore this for
the moment. The usual argument then goes as follows. To leading order, Hawking
radiation may be computed by substituting the classical solution
(corresponding to the formation of a blackhole due to an incoming matter shock
wave along $\s^+ =\s^+_0$) into the
 quantum piece of the stress tensor $T^f$, and then imposing boundary
conditions. In terms of the asymptotically Minkowski coordinates
$\bar\s^+={1\over\l}\log (\l\s^+),~\bar \s^-=-{1\over\l}\log
(-\l\s^--{a\over\l}),$ the classical solution is
 $$\eqalign{2\r&=-\log (1+{a\over\l}e^{\l\s^-}),~~~~~~~~~~~~~\s^+ <\s^+_0,\cr
              2\r&=-\log (1+{a\over\l}e^{\l
(\s^--\s^++\s^+_0)}),~~\s^+>\s^+_0.\cr}$$

Substituting this in $T^f_{--}$ and demanding that the latter vanishes for
$\s^+<\s^+_0$  one determines
$t_-(\s^-)=-{\l^2\over 4}(1-{1\over (1+{a\over\l}e^{\l\s^-})^2})$. Then
observing that $\pa\r ,\pa^2\r\rightarrow 0$ when $\s^+\rightarrow\infty$ (
${\cal I_R^+}$ in the Penrose diagram) we have
$$T^f_{--}\rightarrow {N\over 24}\l^2(1-{1\over (1+{a\over\l}e^{\l\s^-})^2}.$$

This determines the Hawking radiation rate at time like future infinity to be
${N\over 24}\l^2$ in agreement with earlier calculations (see for instance
[\bd ]).

This calculation however neglects back reaction. This is of course true  for
all
previous calculations of Hawking radiation. In the original calculations [\hawk
,\bd ] one quantized in a fixed background metric which means  that back
reaction is ignored. But there is no sensible approximation in
which back reaction can be ignored. Back reaction is of the same order as the
radiation! Within the context of this toy model and our explicit solution of
it, this problem can be resolved. But before we do it let us elaborate on this
question further.

The point is that the one loop (matter) corrected theory has an action
(equation (23) of [\call ] ) and associated equations of motion and
constraints. Aside from the dilaton equation, these correspond to \conf\ and
\cons\ and read in this notation,
$$\eqalign{T_{-+}&=T^{cl}_{-+}+T^f_{-+}=0,\cr
           T_{\mp\mp}&=T^{cl}_{\mp\mp}+T^f_{\mp\mp}=0.\cr}\eqn\teqn$$
One has to now find a consistent solution to this set of equations (and the
$\phi$ equation of motion). Such a solution will have a classical piece plus a
one loop quantum correction. Now in calculating $T^f$
	to order $\hbar$ it is
sufficient to substitute the classical part of the solution into it. But to the
same order one should keep the result of substituting the $O(\hbar )$
correction to the classical solution into $T^{cl}$. One should not just  keep
the former as Hawking radiation and ignore the latter. In fact the classical
solution, by definition, satisfies the classical equations
$T^{cl}_{-+}=0,~T^{cl}_{\pm\pm}=0$, so that in order to satisfy \teqn, the
leading quantum correction to the classical solution when substituted into
$T^{cl}$ must give a value which exactly cancels the value obtained by
substituting the classical solution into $T^f$. The CGHS calculation of course
agrees with the calculations involving quantization in a fixed background,
since
keeping the background fixed is tantamount to ignoring the quantum correction
to the classical solution, and is of course inconsistent with the  quantum
corrected equations of motion and  constraint.

A related point is that the energy-momentum conservation equation and the
equation of motion for $\r$ make $T_{\pm\pm}$ chiral fields as in conformal
field theory.\foot{Indeed the theory is, as we argued earlier, a conformal
field theory.} This is because in any conformal gauge the stress tensor
conservation law (which is a consequence of general covariance and the
matter-dilaton equations of motion) takes the form
$$\pa_{\pm}T_{\mp\mp}+\pa_{\mp}T_{+-}-2\pa_{\mp}\r T_{+-}=0$$

and the equation of motion for $\r$ is equivalent to the first equation of
\teqn, so that $\pa_{\pm}T_{\mp\mp}=0$. Since this is automatically true for
$t$ it is also true separately for  the non-ghost part of the stress tensor.
Now how can we identify the "radiation" part of the stress tensor. As we argued
earlier it does not make sense to just subtract off the "classical" part of the
stress tensor. One can subtract the classical value of the classical stress
tensor (i.e. the value when the classical solution  is substituted into it).
But by definition this is zero, so we are left with the whole stress tensor.
Also  as we've  seen, $T_{--}$ is independent of $\s^+$, and hence cannot be
zero
in the
region $\s^+ <\s^+_0$, and non-zero for $\s^+>\s^+_0$. Indeed since $T$ in this
section is defined to include the ghost contribution (the translation
is $-{N\over 6}t_{\pm}\rightarrow t_{\pm\pm}$)  it is zero everywhere, for that
is the equation of constraint (second equation of \teqn ).

\chapter{\bf A Proposal for Calculating Hawking Radiation}

 How then can we identify Hawking radiation?
In general relativity there is a definition of the energy left in a system
which is asymptotically flat, after radiation has been emitted for a certain
time. This is the so-called Bondi mass. This is defined relative to some
reference static solution and must be given in asymptotically Minkowski
coordinates. So if $\d T_{\mu\nu}$ is the first variation of the
stress tensor around the reference solution, then for a solution (static or
non-static)  which asymptotically approaches the static solution at future
 null infinity, the Bondi mass is given as ($\bar\s^{\pm}$ are the
asymptotically
Minkowski coordinates)
$$M(\bar\s^-)=\int^{\cal I^+_R}d\bar\s^+\d T_+^{~0}=-\int^{\cal
I^+_R}d\bar\s^+(\d T_{++}+\d T{+-}).\eqn\mass$$
 In the above the integral is to be evaluated at the  future null infinity line
${\cal I^+_R}$, i.e. at $\bar\s^+\rightarrow\infty$. Now the linearized stress
tensor satisfies the linearized conservation equation

 $$\pa_{\mp}\d T_{\pm\pm}+\pa_{\pm}\d T_{+-}=0.\eqn\lin$$

 Using this we find from \mass ,
 $$\eqalign{\pa_-M(\bar\s^-)&=-\int^{\cal I^+_R}d\bar\s^+(\pa_-\d
T_{++}+\pa_-\d T_{+-})\cr
&=+\int^{\cal I^+_R}d\bar\s^+(\pa_+\d T_{+-}+\pa_+\d T_{--})\cr
&=(\d T_{+-}+\d T_{--})_{\cal I^+_R}\cr}\eqn\rate$$

This equation gives the rate of decay of the Bondi mass. We may therefore
identify the negative of the right hand side as the radiation flowing out to
future null infinity.

To proceed we need the exact solutions of our quantum corrected equations of
motion (4.3) or (4.4). Once a coordinate system is chosen, these
solutions are given in terms of two unknown chiral functions
$u_{\pm}(\s^{\pm})$ which need to be determined from the constraint equations
and the boundary conditions. As we argued in the last section the boundary
conditions that have been  used in the past, do not make sense because of the
chirality of
the stress tensor, so we have to proceed in an alternative manner. Let us first
impose the constraint equations \cons .

The stress tensor calculated from \newaction\ is

 $$\eqalign{T_{\pm\pm}=&{1\over
2}(\pa_{\pm}X\pa_{\pm}X-\pa_{\pm}Y\pa_{\pm}Y)+\sqrt{|\k |\over 2}\pa^2_{\pm}Y
 +{1\over 2}\sum_i\pa_{\pm}f^i\pa_{\pm}f^i\cr
 =&e^{-2\phi}(4\pa_{\pm}\phi\pa_{\pm}\r-2\pa^2_{\pm}\phi +O(\k
e^{2\phi}))+{1\over
2}\sum_i\pa_{\pm}f^i\pa_{\pm}f^i+\k(\pa_{\pm}\r\pa_{\pm}\r-\pa^2_{\pm}\r
),\cr}\eqn\stress$$
and

$$T_{+-}=-\sqrt{\k\over
2}\pa_+\pa_-Y-\l^2e^{\sqrt{2\over |\k |}(X+Y)}.\eqn\stresstrace$$

In the coordinate system in which $g_{\pm}$ are zero we have from
 (4.3),
$$\eqalign{T_{\pm\pm}=&{1\over 2}\sum_i\pa_{\pm}f^i\pa_{\pm}f^i+
\sqrt{\k\over2}\pa_{\pm}^2Y\cr=&{1\over
2}\sum_i\pa_{\pm}f^i\pa_{\pm}f^i+\pa_{\pm}^2u_{\pm}.\cr}$$
Hence the constraint equations \cons\ become,

$$\pa_{\pm}^2u_{\pm}+{1\over
2}\sum_i\pa_{\pm}f^i\pa_{\pm}f^i+t_{\pm\pm}=0.\eqn\ueqn$$

Now we have the problem of determining  the ghost stress tensor $t$. This, as
well as
the non-ghost stress tensors $T^{X,Y},T^f$, transform like  connections under
coordinate
transformation because of the conformal anomaly. It is only the sum which
transforms as a tensor (since the conformal anomalies cancel between the two).
Thus under a conformal coordinate transformation $\s^{\pm}\rightarrow\s
'^{\pm}=f^{\pm}(\s^{\pm})$,

$$\eqalign{T'^f_{\pm\pm}(\s ' )=&\left ({\pa f^{\pm}\over\pa\s^{\pm}}\right
)^{-2}[T_{\pm\pm}^f(\s )+{N\over 12}Df^{\pm}],\cr
T'^{X,Y}_{\pm\pm}=&\left ({\pa f^{\pm}\over\pa\s^{\pm}}\right
)^{-2}[T_{\pm\pm}^{X,Y}(\s )+{26-N\over 12}Df^{\pm}],\cr
t_{\pm\pm}'(\s ' )=&\left ({\pa f^{\pm}\over\pa\s^{\pm}}\right
)^{-2}[t_{\pm\pm}(\s )+{-26\over 12}Df^{\pm}],\cr}\eqn\trans$$

where $Df$ is the Schwartz derivative  defined by,

$$Df={f'''\over f'}-{3\over 2}\left ({f''\over f'}\right )^2.$$

At this point we do not know how to proceed without making an assumption about
the boundary conditions. We assume that in a preferred coordinate system,
namely
one which covers the whole space (i.e. including the region behind the
classical horizon) and is asymptotically Minkowski, the expectation value of
the  matter stress tensor $T_{--}^f$
vanishes. These coordinates are related to the Kruskal-Szekeres coordinates by

$$\hat\s^+={1\over\l}\log (\l\s^+),~\hat\s^-=-{1\over\l}\log
(-\l\s^-).\eqn\hatcdt$$

This condition seems to correspond to Hawking's boundary condition, and the
reasoning is that there should be no $f$-particle energy coming in from ${\cal
I_L^-}$. Now from the point of view of the exact theory  the total (including
ghosts)  stress tensor is zero, so that it is difficult to see what objective
meaning this condition has. Nevertheless in order to be as close as possible to
the original calculation, let us impose,

$$\hat T^f_{\pm\pm}=0,~~\hat T^{X,Y}_{\pm\pm}+\hat t_{\pm\pm}=0.$$

The latter follows from the first equation and the constraint. However it still
leaves
us the freedom of choosing the separate values of the ghost and $X,Y$ stress
tensors. Let us put $\hat t_{--}=\a{\l^2\over 24}=-\hat T^{X,Y}_{--}$ i.e. we
have in the $\hat\s$ frame, an arbitrary constant influx of ghost stress energy
balanced by a constant outflow of $X,Y$ stress energy. On ${\cal I _R^-}$ there
is incoming $f$ stress energy, which following CGHS [\call ] we take to be
$\hat T^f_{++}=a\l\s^+_0\d (\hat\s-\hat\s_0)$. Then we may take $\hat
t_{++}=\a{\l^2\over 24}$ and $\hat T^{X,Y}_{++}=-a\l\s^+_0\d (\hat s-\hat
s_0)-\a{\l^2\over 24}$ to be consistent with the constraints.

 Then by putting $\s '=\hat\s$ in \trans, we get
in the $\s$ frame,

$$t_{\pm\pm}=-{26-\a\over 24}{1\over\s^{+2}},~T^f_{--}=-{26\over 24}{1\over
\s^{-2}},$$
$$T^f_{++}=-a\l\s^+_0\d (\hat s-\hat s_0)-{N\over 24}$$

Using these values in \ueqn\ we find,

$$\eqalign{u_+=&a_++b_+\s^+-a(\s^+-\s^+_0)\t (\s^+-\s^+_0) -{\bar N\over
24}\log
|\s^+|,\cr u_-=&a_-+b_-\s^--{\bar N\over 24}\log |\s^-|,\cr}\eqn\usoln$$
 where $$\bar N=N+\a-26\eqn\enbar$$.

 We now need a reference static solution. This is obtained from \ssoln\ and
\usoln\ by putting $a=a_{\pm}=b_{\pm}=0$ in the latter;

 $$X_0=-Y_0=\sqrt{2\over |\k |}(\l^2\s^+\s^-+{\bar N\over 24}\log
(-\s^+\s^-)),~
f=0.$$
This solution is in  Kruskal-Szekeres coordinates, and
we need to transform this into the  asymptotically Minkowski
coordinates\foot{For the {\it static} solution these are the same as $\hat\s$
defined above.}
$\bar\s^{\pm}$ defined by  $\s^+={1\over\l}e^{\l\bar\s^+},
\s^-=-{1\over\l}e^{-\l\bar\s^-}$.
Under a coordinate transformation $X$ transforms as a scalar, and (since $\r(\s
)\rightarrow \r(\bar\s )+{\l\over 2}(\bar\s^+-\bar\s^-)$) $Y$ transforms as
$Y(\s )\rightarrow Y(\bar\s )+\sqrt{|\k |\over 2}\l(\bar\s^+-\bar\s^-)$. Hence
we have in the new coordinate system,

$$\eqalign{X_0=&-\sqrt{2\over |\k |}\left (e^{\l(\bar\s^+-\bar\s^-)}-{\bar
N\over
24}\l(\bar\s^+-\bar\s^-)+{\bar N\over 24}\log\l^2\right )\cr
=&-Y_0+\sqrt{|\k |\over 2}\l (\bar\s^+-\bar\s^-).\cr}\eqn\static$$

This solution corresponds to the linear dilaton solution of the classical
equations.
 To obtain the Bondi mass of a general solution which  asymptotically tends to
the above static solution, we need to linearize the stress tensors around the
latter. From \stress\ and \stresstrace\ we have using \static,
 $$\eqalign{\d T_{++}+\d T_{+-} =&-\sqrt{2\over |\k |}\pa_+\left [\l e^{\l
(\bar\s^+-\bar\s^-)}
 (\d X+\d Y)\right ]+\sqrt{2\over |\k |}{\bar N\over 24}\l\pa_+ (\d X+\d
Y)\cr&-\sqrt{|\k |\over 2}\l\pa_+\d Y+\sqrt{|\k |\over 2}(\pa_+(\pa_+\d
Y-\pa_-\d Y)\cr.}$$

Substituting into \mass\ we get

$$\eqalign{M(\bar\s^-)= & -\int d\bar\s^+(\d T_{++}+\d T_{+-})\cr
= & \Bigg [\sqrt{2\over \k}\l e^{(\bar\s^+-\bar\s^-)}(\d X+\d Y)
-\sqrt{2\over |\k |}{\bar N\over 24}\l (\d X+\d Y)\cr & +\sqrt{|\k |\over
2}\l\d Y-
\sqrt{|\k |\over 2}(\pa_+\d Y-\pa_-\d Y)\Bigg ]_{\cal I^+_R}.\cr}\eqn\bondi $$

Using \xcdt\ and\ycdt\ we find that when $e^{2\phi}<<1$ this expression tends
(not surprisingly) to the expression given by CGHS (equation (26) of [\call ])
except for the ghost terms.

Static solutions corresponding to black holes (in the classical limit) are
obtained by putting $a=b_{\pm}=0$ and $a_{\pm}\ne 0$. Then for $\bar\s^+>>1$,
$-\d Y=\sqrt{2\over \k}(a_++a_-)=\d X$; and we have from \bondi, a constant
Bondi (ADM) mass
$$M(\bar\s^-)= \l (a_++a_-).$$

The parameters $a_{\pm}$ can be of either sign and hence we may have negative
mass solutions of the theory. Of course the classical theory has such solutions
too,
but there these correspond to naked singularities, whereas here these are
non-singular solutions (as we argued in section 4)\foot{This point has been
emphasized  by Giddings and Strominger [\gs ].}. However one might ask whether
it is the case that  we cannot  generate  these  unphysical solutions
dynamically, by starting with positive mass solutions, in which case we might
choose to ignore them. Unfortunately this is not the case. To see this, let us
compute the Bondi mass of the analog of the dynamic CGHS solution corresponding
to the formation of a black hole by an incoming matter shock wave, and its
decay
by Hawking radiation. This solution is obtained in the $\s$ frame by putting
$a_{\pm}=b_{\pm}=0,~a\ne 0,$ in \ueqn\ and substituting in \ssoln. Then in the
region outside the classical horizon we transform to the asymptotically
Minkowski coordinates $\s$ defined by,
$$\s^+={1\over\l}e^{\l\bar\s^+},~~\s^-=
-{1\over\l}e^{-\l\bar\s^+}-{a\over\l^2},$$
 to get,
 $$\eqalign{X =& -\sqrt{2\over |\k |}\left [{M_0\over\l}\t
(\bar\s^+-\bar\s^+_0)+e^{\l (\bar\s^+-\bar\s^-)}-{\bar N\over 24}\log \left
({e^{\l\bar\s^+}\over\l}\left ({e^{-\l\bar\s^-}\over\l}+{a\over\l^2})\right
)\right ) \right ]\cr
  =& -Y+\sqrt{\k\over2}\l (\bar\s^+-\bar\s^-),\cr}$$
where we have put $M_0=\l a\s^+_0$ the mass of the classical black hole.
Comparing with the static solution we find,

$$\d X=\d Y=-\sqrt{2\over |\k |}\left [{M_0\over\l}\t
(\bar\s^+-\bar\s^+_0)-{\bar N\over 24}\log
\left (1+{a\over\l}e^{+\l\bar\s^-}\right )\right ].$$

Substituting into \bondi\ we get,

  $$M(\bar\s^- )=M_0-{\bar N\over 24}\l\log (1+{a\over\l}e^{\l\bar\s^-})-{\bar
N\over
24}{\l\over 1+{\l\over a}e^{-\l\bar\s^-}}.$$

 In the infinite (light cone time) past $\bar\s^-\rightarrow -\infty$ the Bondi
mass tends to the classical black hole mass $M_0$, but at future infinity
$\bar\s^-\rightarrow +\infty$ one gets an infinitely negative value.

 This unphysical conclusion is equivalent to the statement that the Hawking
 radiation rate does not go to zero asymptotically. This rate may be calculated
either from the left hand side, or as the negative of the right hand side, of
\rate\
$$-{d M(\bar\s^-)\over d\bar\s^-}={\bar N\over 24}{\l^2\over (1+{\l\over
a}e^{-\l\bar\s^-})^2}
\rightarrow {\bar N\over 24}\l^2.\eqn\rad $$

Now so far $\a$ has been kept arbitrary, but perhaps the most natural choice is
$\a = 26$, so that (see \enbar ) $\bar N= N$. This choice corresponds to the
decoupling of the ghosts from the   Hawking  radiation  which is positive
regardless of the number of matter fields. This agrees with the two dimensional
analog of the  original Hawking result [\hawk ],   as well as that of [\call ]
asymptotically but back reaction still modifies the $\s^-$ dependence.
Unfortunately although the formalism allows this value of $\bar N$ it is
certainly not the only possibilty. Perhaps this choice has to made on physical
grounds. It is also possible that our analysis of the Bondi mass is not the
complete quantum mechanical story, and a proper treatment would resolve both
this issue as well as the question of positivity.
\chapter{\bf Conclusions}

What progress have we made in understanding quantum black holes, and in
particular the phenomenon of Hawking radiation, from this work?

Firstly let us stress that even if we leave aside our argument for regarding
the Liouville-like theory as the complete quantum theory, it still gives us
the only consistent treatment of the semi-classical theory (i.e. to first order
in $\k e^{2\phi}$). As we pointed out in section three, if we just include the
leading order corrections then the field space curvature is zero, and one
immediately has a soluble semi-classical theory. All of the above calculations
are then still
valid except that we cannot draw some of the conclusions that we've drawn from
them. Thus we can no longer explicitly demonstrate the taming of the classical
singularity, and of course there is no need to conclude that the Hawking
radiation does not stop, and that positive mass black holes radiate into
negative mass solutions. Nevertheless one has a consistent  semi-classical
 picture of black hole radiation and back reaction. In particular it should be
emphasized again that our remarks about the  inconsistencies associated with
the usual calculation of Hawking radiation which ignores back reaction, are
valid already at the semi-classical level. To belabor the point, the
calculations
with the Liouville-like theory, when interpreted in terms of the $\r ,\phi$
variables, and  considered as being valid to $O (\k e^{2\phi})$, are the
correct semi-classical results coming from the classical CGHS theory. In
particular, they  show that the same semi-classical physics is obtained
whatever options are chosen for the functions $h,\bar h$ (as discussed in
section three) simply because $\bar{\bar h}$ is zero at the semi-classical
level. In other words we may use the exactly soluble conformal field theory to
make the calculations, provided we interpret the result as being valid only at
the semi-classical level.

Secondly we have shown that there is a class of quantum dilaton gravity
theories,
namely those for which the field-space curvature is zero to all orders in
$e^{2\phi}$ ($\bar{\bar h}=0$, see
discussion after equation (3.1)) whose exact quantum treatment is possible
since they can be transformed into a Liouville-like theory.\foot{The objection
raised by some authors on the range of integration has been  answered in
section 3. In particular for the sub-class d) there can be no objection on
these grounds.} These theories allow for the first time a complete quantum
mechanical treatment (including the effects of dilaton-graviton loops) of a
theory of gravity with classical black hole solutions. Unfortunately, as we
have shown, these theories may  not be physical. It is an open question whether
it
is possible to find a soluble theory with $\bar{\bar h}\ne 0$ which does not
have this problem, but we believe that this is unlikely. After all as we have
explicitly demonstrated, quantum mechanics does what one expects it to do,
namely it tames the classical singularities, including the naked ones!
However it thereby eliminates the usual argument (in the classical
theory) for eliminating negative mass solutions on the grounds that such spaces
are not globally hyperbolic.   It is possible that the problem is not so much
with the soluble class of models that we have treated, as with the original
classical dilaton-gravity theory itself, which  does not have a positive
definite field space metric. On the other hand it is also possible that the
fault lies with our rather heuristic treatment of the Bondi mass in the quantum
theory, and that a rigorous quantum treatment may resolve this issue.

\chapter{\bf Acknowledgements}

I wish to thank Geoff Harvey, Greg Moore, and Andy Strominger, for discussions.
This work is partially supported by Department of Energy contract No.
DE-FG02-91-ER-40672.

\refout

\end